\documentstyle[12pt,aaspp4]{article}

\newcommand{\nix}{$\cdot\cdot\cdot$}
\def \la{\mathrel{\mathchoice
{\vcenter{\offinterlineskip\halign{\hfil
$\displaystyle##$\hfil\cr<\cr\sim\cr}}}
{\vcenter{\offinterlineskip\halign{\hfil$\textstyle##$\hfil\cr
<\cr\sim\cr}}}
{\vcenter{\offinterlineskip\halign{\hfil$\scriptstyle##$\hfil\cr
<\cr\sim\cr}}}
{\vcenter{\offinterlineskip\halign{\hfil$\scriptscriptstyle##$\hfil\cr
<\cr\sim\cr}}}}}
\def \ga{\mathrel{\mathchoice
{\vcenter{\offinterlineskip\halign{\hfil
$\displaystyle##$\hfil\cr>\cr\sim\cr}}}
{\vcenter{\offinterlineskip\halign{\hfil$\textstyle##$\hfil\cr
>\cr\sim\cr}}}
{\vcenter{\offinterlineskip\halign{\hfil$\scriptstyle##$\hfil\cr
>\cr\sim\cr}}}
{\vcenter{\offinterlineskip\halign{\hfil$\scriptscriptstyle##$\hfil\cr
>\cr\sim\cr}}}}}

\slugcomment{To Appear in the Astrophysical Journal (Main Journal)}

\lefthead{Weaver et al.}
\righthead{X-ray Emission from NGC 1052}

\begin{document}

\title{X-ray Emission from the Prototypical LINER Galaxy NGC 1052}

\author{K. A. Weaver\altaffilmark{1}}
\affil{Johns Hopkins University, Department of Physics and Astronomy,
Homewood campus, 3400 North Charles Street, Baltimore, MD
21218; kweaver@pha.jhu.edu}

\author{A. S. Wilson\altaffilmark{2}}
\affil{Space Telescope Science Institute, 3700 San Martin Drive,
Baltimore, MD 21218; awilson@stsci.edu}

\author{C. Henkel}
\affil{Max-Planck-Institut f{\"u}r Radioastronomie,
Auf dem H{\"u}gel 69, D-53121 Bonn, Germany; p220hen@mpifr-bonn.mpg.de}

\and

\author{J. A. Braatz}
\affil{National Radio Astronomy Observatory, P. O. Box 2, Green Bank,
WV 24944; jbraatz@nrao.edu}

\altaffiltext{1}{Also NASA Goddard Space Flight Center, 
Greenbelt, MD 20771}

\altaffiltext{2}{Also Astronomy Department, University of Maryland,
College Park, MD 20742; wilson@astro.umd.edu}

\begin{abstract}

We examine the 0.1 to 10.0 keV X-ray spectrum of the 
bright nuclear LINER galaxy NGC 1052,
one of two elliptical galaxies known to contain a luminous H$_2$O
maser.  The observed $2.0-10.0$ keV spectrum is unusually flat
(photon index $\Gamma\sim0.2$) and best described as
intrinsically power-law shaped nuclear flux that is either (1)
attenuated by a complex absorber with $\sim70\%$ of the nuclear 
flux absorbed by a column density of 
$N_{\rm H} \sim 3\times10^{23}$
cm$^{-2}$ and $\sim30\%$ absorbed by a column
density of $N_{\rm H} \sim3-5\times10^{22}$ cm$^{-2}$, or (2)
reprocessed, with the nuclear source blocked and the X-rays 
Compton reflected in our direction by high column density 
($\ge10^{24}$ cm$^{-2}$) gas.  The moderate equivalent
width of the Fe K$\alpha$ line favors the dual 
absorption model as the most likely
scenario.  The $0.1-2.0$ keV spectrum does not resemble the few
times 10$^6$ to 10$^7$ K thermal emission typically found in
other elliptical galaxies, but instead is best described as 
nuclear X-rays leaking through a patchy absorber or scattered 
in our direction by low-density, ionized gas with the thermal
contribution limited to about $15\%$ for solar abundances.   
The absorption-corrected $2-10$ keV luminosity
of the nuclear source is L$_X \sim 8\times10^{41}$ ergs s$^{-1}$
or L$_X \sim 2\times10^{43}$ ergs s$^{-1}$ for the
dual-absorption and Compton-reflection models, respectively.
The absorbing and H$_2$O masing gases appear to be spatially 
separate in this galaxy.

\end{abstract}


\newpage
\section{Introduction}

Water vapor ``megamasers'' with isotropic luminosities of $\sim
20$ to 6000 L$_{\odot}$ have been detected in eighteen galaxies
to date (Braatz, Wilson \& Henkel 1996; Greenhill et al.\ 1997;
Hagiwara et al.\ 1997).  All of the megamaser galaxies show 
evidence for nuclear activity, being classified as either Seyfert
2 or LINER, and so there appears to be a link between the 
megamaser phenomenon and AGN activity.  However, no megamasers
have been detected in Seyfert 1 galaxies (Braatz, Wilson, \&
Henkel 1997).  This suggests that either Seyfert 1s do not possess
enough molecular gas with the physical conditions appropriate 
to produce 1.3 cm H$_2$O masers or the masers are beamed away 
from Earth.

There is strong evidence that H$_2$O megamasers are associated 
with the dense molecular gas located at the centers of galaxies.
Interferometric studies show that the maser emission originates
within about 1 pc or less of the nucleus (Claussen \& Lo 1996;
Miyoshi et al.\ 1995; Claussen et al.\ 1986).  In particular,
observations of maser emission from gas rotating at high
velocities near the center of the LINER galaxy NGC 4258 indicate
the presence of a thin, nearly edge-on molecular disk having an
inner radius 0.13 pc, an outer radius 0.25 pc, and a significant
warp (Miyoshi et al.\ 1995).  These observations supply perhaps
the best direct evidence to date for a circumnuclear disk of 
accreting matter in an active galaxy, and imply the presence of
a $3.6 \times 10^7$ M$_{\odot}$ black hole at the center of 
NGC 4258.

The nature of H$_2$O megamasers and their relation to AGNs
are still not clear.  It may be that megamasers are present 
in a large fraction of AGNs, but are detected only in those 
for which the nuclear disk (a ubiquitous feature in unified 
models) is viewed close to edge-on.  Alternatively,
megamasers may result from special geometrical and/or 
physical conditions.  For example, in the H$_2$O maser-detected
galaxy NGC 4258 there is a combination of a nuclear hard X-ray 
source and a thin, warped molecular disk.  The hard X-ray 
source irradiates the warped disk and gives rise to a 
molecular layer at temperatures $400-1000$ K
within which the water abundance is large 
(Neufeld \& Maloney 1995).  According to Neufeld, Maloney
\& Conger (1994), the physical conditions within this layer 
give rise to collisionally pumped H$_2$O maser emission
with a predicted luminosity of $10^{2\pm0.5} L_{\odot}$
per pc$^2$ of illuminated area.  Another possible geometry
involves a nuclear hard X-ray source surrounded by a 
geometrically thick molecular torus, with the maser emission 
arising from the inner surface of the torus (e.g., Greenhill
et al.\ 1996).  Of course, other sources of energy (e.g., 
shock waves, cf. Maoz \& McKee 1998) may also be significant 
in heating the molecular gas to temperatures at which 
collisional pumping is effective.  Nevertheless, the likelihood
that X-ray heating is important and the tendency for H$_2$O 
megamaser-detected galaxies to have high column densities
to their nuclei (Braatz, Wilson \& Henkel 1997) indicates 
that X-ray observations are crucial for understanding the 
water vapor megamaser phenomenon.

To address these questions, we have undertaken a study of 
megamaser galaxies with the Advanced Satellite for Cosmology
and Astrophysics (ASCA; Tanaka, Inoue \& Holt 1994).  Here 
we present results for the bright nuclear LINER
NGC 1052 (z = 0.0049, D = 19.6 Mpc
for H$_0$ = 75 km s$^{-1}$ Mpc$^{-1}$; 
D = 29.4 Mpc for H$_0$ = 50 km s$^{-1}$ Mpc$^{-1}$) which has 
$L_{maser} {{\sim}\atop{=}} 200 L_{\odot}$ (Braatz, Wilson \& Henkel 1994).
NGC 1052 is one of seven megamaser galaxies classified as 
LINERs and one of only two that are ellipticals.  The 
spectral profiles of the masers in elliptical galaxies
differ from the features found in other active nuclei.
The emission lines are broad with $\sim90$ km s$^{-1}$
FWHM and are not resolved into narrow components.  VLBI
observations of NGC 1052 show that the masers are not 
located in a disk around the central engine as in NGC 4258,
but rather lie along the direction of the radio jet
(Claussen et al.\ 1998).  NGC 1052 was detected with the
$Rosat$ PSPC with a $0.1-2.0$ keV flux of $\sim6\times10^{-13}$
ergs cm$^{-2}$ s$^{-1}$ (Brinkmann, Siebert \& Boller 1994).
The $ASCA$ observation is the first in the $2-10$ keV band.

\section{Observations and Analysis}

NGC 1052 was observed with $ASCA$ on 11 August 1996.  $ASCA$ 
contains four imaging telescopes that focus X-rays onto two 
pairs of detectors located at the focal plane.  The detectors 
are the solid state imaging spectrometers (SIS; two detectors 
labeled S0 and S1) and the gas imaging spectrometers (GIS;
two detectors labeled G2 and G3).  The SIS data were obtained 
in 1-CCD readout mode, in which one CCD chip was exposed on each 
SIS, and were converted from FAINT to BRIGHT telemetry mode 
for analysis.  The GIS data were obtained in 
standard pulse-height mode.  

To produce good time intervals, the $ASCA$ data were screened 
according to the following criteria.  For all detectors,
data collected during the satellite's passage through the 
South Atlantic Anomaly (SAA) were discarded, as were data 
collected when satellite elevation angles were $\le5^{\circ}$.
For the SIS, additional data were discarded when the 
geomagnetic cut-off rigidity (COR) was $\le6$ GeV c$^{-1}$,
when elevation angles during times of bright Earth were 
$\le20^{\circ}$, for times $\le16$ s after an SAA passage,
and for times $\le16$ s after a transition from satellite 
day to satellite night.  A prominent spike that remained 
in the SIS background after screening was removed, hot and 
flickering pixels were removed, and grade 0,2,3 and 4 photon 
events were selected.  For the GIS, rise-time rejection was
applied to exclude particle events, data were discarded for 
COR$\le4$ GeV c$^{-1}$, and times of soft and hard background flares 
were rejected (as described in the ASCA Data Reduction 
Guide 1997).  After screening, 37.2 ks of good data were 
collected from each GIS and 37.6 and 36.2 ks of good data 
were collected from S0 and S1, respectively.

Spectra were extracted from each detector and grouped to have 
at least 20 counts per bin to allow the use of $\chi^2$
statistics.  Background was obtained from source-free regions 
in the same field.  Our results do not change when 
using background from the standard blank-sky fields.
The background-subtracted count rates for 
NGC 1052 range from 0.040 to 0.049 counts s$^{-1}$ per
detector, with the 
background accounting for $20\%$ and $30\%$ of the total rate in 
the SIS and GIS, respectively.  Detector response matrices 
were generated with version 4.0 of the FTOOLS software.  The 
four spectra were fitted together with version 10.0 of the 
XSPEC spectral fitting package (Shafer, Haberl \& Arnaud
1989) and the model normalizations for each dataset were 
allowed to vary independently.

The $Rosat$ PSPC observed NGC 1052 
on 23 July 1993 for approximately 14 ks.  The galaxy was
also observed for a total of about 25 ks with the HRI,
but since we focus on the spectrum,
we do not include the HRI data in our analysis
(see Guainazzi and Antonelli 1998).  The PSPC spectrum 
was extracted from a circular region in the source field
of about 1.2 arcmin radius.  Background was taken from 
source-free positions located about 4 arcmin away from 
and surrounding the galaxy.  Subtracting background yields
a PSPC source count rate of 0.035 counts s$^{-1}$.

\section{Results of Spectral Fitting}

The observed $0.1-2.0$, $0.6-2.0$ and $2-10$ keV luminosities of NGC 1052
are $\sim 4\times10^{40}$ ergs s$^{-1}$, $\sim 3.4\times10^{40}$ ergs s$^{-1}$,
and $\sim 4\times10^{41}$ ergs s$^{-1}$, respectively (all luminosites
are quoted for H$_0$ = 50 km s$^{-1}$ Mpc$^{-1}$).  Figure 1 shows the 
background-subtracted $ASCA$ spectrum.  The spectrum has been 
adapted for illustration purposes to have the best possible 
ratio of signal to noise at all energies.  The data were 
averaged according to detector type, coarsely binned, and then 
plotted in the energy band for which that detector type has 
superior efficiency: the SIS below 5 keV and the GIS above 
5 keV.  Instrumental effects have not been removed, but can 
be estimated from the plotted telescope-plus-detector
effective area curve. 

The complex shape and hardness of the spectrum rules out a 
thermal bremsstrahlung or an ionized thermal plasma origin for
the X-ray emission.  Applying such models to the $ASCA$ plus
PSPC data yields extremely poor fits ($\chi^2/\nu = 2.8$) and 
implies unphysically high temperatures of kT $>$ 100 keV.  
Restricting the fits to energies less than 2.0 keV yields 
a temperature of kT $>$ 3 keV, which is consistent with 
that measured by Davis and White (1996)
and is much higher then the temperatures of other elliptical
galaxies with velocity dispersions similar to that of NGC 1052.   
In fact, Davis and White (1996) discard NGC 1052 from 
their statistical analysis of X-ray properties of elliptical 
galaxies based on the argument that the soft X-ray flux is 
dominated by X-ray binaries or an AGN.  Evidence for 
extended emission in the HRI (Guainazzi \& Antonelli 1999)   
indicates some contribution from the galaxy which we 
investigate in \S 3.4. 

\subsection{Simple AGN Models}

The hardness of the spectrum between 2 and 10 keV and the visible 
Fe K$\alpha$ fluorescence line at 6.4 keV verify the assertion 
made by Davis \& White (1996) that the X-ray spectrum of NGC 1052 
is dominated by something other than diffuse X-ray emitting gas,
most likely an AGN.  We favor this idea as opposed to a `normal'
ensemble of X-ray binaries for the following reason. 
The observed correlation between 
hard X-ray luminosity (L$_{X}$) and 
absolute blue-band (L$_{B}$) luminosity for early-type galaxies
(Matsumoto, et al.\ 1997; Matsushita et al.\ 1994) predicts 
that X-ray binaries 
should contribute $\la 10^{40}$ ergs s$^{-1}$ for NGC 1052  
(log L$_{B}$/L$_{\odot}$ = 10.2); however this represents 
$\la2\%$ of the observed hard X-ray luminosity.

Unobscured AGN have approximately power-law
X-ray spectra in the $ASCA$ band and so we make the simplifying 
assumption that the {\it intrinsic} $0.6-10.0$ keV spectrum has 
a power-law shape.  We first consider simple AGN continuum models 
of the following form: (1) a power law with uniform, neutral 
absorption, (2) a power law that is partially covered by a patchy,
neutral absorber, and (3) a uniformly absorbed power law plus flux 
scattered by an ionized electron scattering medium (a warm ``mirror'').
A Gaussian function is added, where appropriate, to model the 
Fe K$\alpha$ line.  We list the best-fitting continuum and line 
parameters in Tables 1 and 2, respectively.
A model that consists of a power law absorbed by the Galactic 
column density of $N_{\rm H} = 3\times10^{20}$ cm$^{-2}$ (Table 1,
model 1) provides a formally unacceptable fit ($\chi^2/\nu = 1.58$
for 325 d.o.f.) and yields an unusually small photon index of 
$\Gamma = 0.24$.  The data divided by this model are shown in 
Figure 2a.  Such a flat power law does a good job of approximating 
the data between 1 and 4 keV, but the data fall systematically 
above and below the model at energies $<1$ keV and $>7$ keV,
respectively.  The model does not include the Fe K$\alpha$
line, which is apparent in the data near 6.4 keV.

The simple power-law continuum model especially fails to 
explain the soft X-ray 
spectrum and so we investigate ways to model the extra flux below 
1 keV (the ``soft excess'').  Our preliminary fits with pure
thermal models (\S 3) allow us to rule out an origin of the soft 
X-rays entirely in a shock-heated plasma such as a galactic wind,
with expected temperature of a few tenths of keV, or in a 
photo-ionized plasma, such as the warm electron scattering mirror,
with expected temperature of $\sim 0.9$ keV, as seen in NGC 1068
(e.g., Marshall et al.\ 1993).  Non-thermal models provide more 
reasonable results.  The soft excess can be described by leakage
of nuclear continuum photons through a patchy absorber (model 2)
or as scattered nuclear X-rays (model 3), but both models still imply 
a photon index of $\Gamma \sim 1.3$, which is smaller than 
the ``canonical'' value for Seyfert galaxies of $\sim1.7$ to 
$\sim1.9$ (Turner \& Pounds 1989; Nandra \& Pounds 1994).  For 
the scattering interpretation, the scattered fraction is $\sim22\%$,
which is much higher than typical values of $\sim 1-2\%$ for 
Seyfert 2 galaxies (e.g., Mulchaey et al.\ 1993), although this 
problem is not as severe for more complex models ($\S\S 3.2-3.4$).

To model the Fe K$\alpha$ line, we 
add an unresolved Gaussian to the power-law 
plus scattered flux model, which improves $\chi^2$ by 20 
(model $3_{\rm G}$).  This represents a significant 
improvement in the fit ($>>99\%$ confidence 
for the addition of two free parameters); however, 
even with such a large improvement, 
positive residuals are seen in all detectors at up 
to the 30\% level near 0.6 and 2 keV (Figure 2b).  Adding 
a second Gaussian to model the curvature in the spectrum 
between 4 and 8 keV allows the photon
to increase and $N_{\rm H}$ to decrease, which in turn 
results in a better fit to the soft excess and $\sim2$ keV bump
and thus improves the fit further (model $3_{\rm 2G}$).
The problem with this model is that   
the line width increases dramatically to the implausible 
value of $\sigma = 2.4$ keV.
This excellent {\it empirical} fit indicates 
more complexity in the spectrum than the superposition of a 
single-absorbed power law and soft X-ray component.  Below 
we investigate the likely possibilities.

\subsection{A Dual-Absorber Model}

Since the flatness of the spectrum between 1.5 and 6 keV
(Figure 1, 2a,b) cannot be described by simple power-law models
(models 1 through $3_{\rm G}$, Table 1),
we investigate physical models that can produce a flat, observed 
spectrum but allow the intrinsic spectrum to be steeper 
and more similar to that expected for an AGN.  One possibility
is that the spectrum is modified by a complex absorption structure.
We therefore tried a continuum model that consists of an 
intrinsically power-law nuclear spectrum of which a fraction 
$C$ is attenuated by a column density $N_{\rm H1}$, a fraction 
$1-C$ is attenuated by a column density $N_{\rm H2}$, and a 
fraction is scattered into our line of sight (the soft excess),
plus a single Gaussian representation of Fe K$\alpha$.  This
so-called ``dual absorber'' model (model 4) is more physically reasonable 
than the double-Gaussian model ($3_{\rm 2G}$) and can explain 
the unusually flat X-ray spectra of other heavily absorbed
AGN such as NGC 4151 (Weaver et al.\ 1994) and NGC 5252 (Cappi
et al.\ 1996).  The corresponding physical scenario might be 
an extended X-ray source, such as an accretion disk corona,
covered by a clumpy or layered absorber.

The dual-absorber model describes the 
data as well as the empirical double-Gaussian model.  
The ratio of power-law normalizations at 1 keV for the 
scattered component ($1.5 \times 10^{-4}$ photons keV$^{-1}$
cm$^{-2}$ s$^{-1}$) and the heavily 
absorbed component ($13 \times 10^{-4}$ photons keV$^{-1}$
cm$^{-2}$ s$^{-1}$) yields a scattered 
fraction of $\sim11\%$.  To fully reproduce the spectrum
we require $70\%$ of the 
intrinsic flux to be absorbed by a large column of 
$N_{\rm H1}=3\times10^{23}$ cm$^{-2}$ and $30\%$ to be 
absorbed by a smaller column of $N_{\rm H2}=5\times10^{22}$
cm$^{-2}$.  
For this model, the absorption-corrected $2-10$ keV flux is 
$8\times10^{-12}$ ergs cm$^{-2}$ s$^{-1}$, yielding 
an intrinsic $2-10$ keV luminosity of $8\times10^{41}$ ergs
s$^{-1}$.

\subsection{Models That Include Compton Reflection}

Another way to produce a flat, hard spectrum (with photon 
index comparable to the observed value of $\Gamma=0.16$ above 
1 keV) is by Compton reflection from a dense, neutral medium
(Lightman \& White 1988; Guilbert \& Rees 1988).  We include a
reflection component in the next set of models by means of the 
{\it hrefl} model in XSPEC.  This model uses an elastic 
scattering approximation that is valid up to energies of 
15 keV and predicts the reflected spectrum from a cold, optically
thick disk illuminated by a point source located at height H 
above the center of the disk. 
The reflected component suffers heavy photoelectric 
absorption, and the model is the sum of X-rays from the 
directly-viewed point source and those reflected from the disk.
The model parameters that can be varied are the height of the 
point source above the disk (here taken to be 
small compared to the disk radius), the abundance 
($Z$) of the gas in the disk, the inclination ($\theta$) of 
the disk normal with respect to the observer's line of sight,
the fraction ($f_i$) of the spectrum of the point source
(here taken to be a power law with index $\Gamma$) that the 
observer sees directly, and a measure ($R$) of the relative 
amount of reflected flux, which is normalized to unity  
when the point source emits isotropically and the 
disk covers $2\pi$ steradians as seen from the 
point source.  If the photon spectrum of the point source is
$N(E)$ and the reflected continuum from the disk in our 
direction is $R(E)$ (both in photons cm$^{-2}$ s$^{-1}$ keV$^{-1}$),
then the model spectrum for a given inclination and abundance
is M(E) = $f_i \times N(E) + R \times R(E)$.  If  
H is $<<$ the radius of the 
disk, the point source emits isotropically, is time stable, 
and is not blocked from our view, then $f_i=1$ and $R=1$.
If, on the other hand, we see none of the X-rays from the 
point source, then $f_i=0$.

Care must be taken when interpreting the reflection model.
The shape of the model X-ray spectrum is a function of $f_i$,
$\Gamma$, and $R \times R(E)$, the last being also
a function of $Z$, $\theta$ and the solid angle subtended by the 
disk at the point source.
Comparison with the observed spectrum can 
provide the relative amounts of directly viewed and reflected 
emission, but interpretation in terms of these parameters is  
ambiguous.  For example, a relatively large amount of reflected 
flux can indicate a large covering factor for the disk 
or a small $f_i$.  The $ASCA$ data cannot determine
$\Gamma$ or $\theta$, so we fix $\Gamma$ at its canonical 
observed value for AGN of 1.7 (Turner \& Pounds 1989), and 
we arbitrarily fix $\theta$ at 60$^{\circ}$.

If the spectrum above 2 keV represented a directly-viewed nuclear
source plus reflection with the canonical parameters $R=1$ and 
$f_i=1$, we would expect the nuclear source to dominate up to 
$\sim8$ keV and then the spectrum would flatten.  Since the 
observed spectrum remains flat down to almost 1 keV, we can  
can immediately rule out this possibility.  
If we instead assume that almost all X-rays above 2 keV 
are reflected (the flattening being due to photoelectric 
absorption), then the fit is acceptable
(Table 1, model 5) and we find $f_i=0.007\pm0.002$ for $R=1$.  
The small $f_i$ implies that the central source is either highly
anisotropic, highly time-variable, or is hidden behind gas with 
a large column density.  While the first two possibilities cannot 
be excluded, we feel that the most plausible explanation for 
the small amount of flux seen directly is that the power-law 
source is blocked from our view.  Assuming full coverage of the 
nuclear X-ray source, we find that the nuclear absorbing column must
be greater than $3\times10^{23}$ cm$^{-2}$ to attenuate the nuclear 
flux so that reflection dominates the spectrum.

One problem for the reflection model is the $\sim40$ eV equivalent 
width (EW) of the Fe K$\alpha$ line, which is much smaller than 
expected for a purely reflected spectrum.  Fully blocked Seyfert 2 
galaxies like Mkn 3 (Iwasawa et al.\ 1994), NGC 4945 (Done, Madejski,
\& Smith 1996), and Circinus (Matt et al.\ 1996) have Fe K$\alpha$
EWs of $\sim 1-2$ keV.  To account for the weak Fe K$\alpha$ line 
within the context of the reflection model, the iron abundance must
be $\le0.05Z_{\odot}$, which is implausibly small.  On the other 
hand, the lower column densities implied by the dual-absorber 
model ($10^{22}$ to a few times $10^{23}$ cm$^{-2}$) allow a 
scenario in which we view the nuclear source directly above 
$\sim5$ keV.  In this case, an Fe K$\alpha$ line with an 
EW of $\sim200$ eV could be produced in a Compton-thick disk  
or in a spherical distribution of clouds with a large covering 
factor.  Since the properties of the Fe K$\alpha$ line  are 
much more consistent with transmission than reflection models,
we argue in favor of the dual-absorber interpretation.

We have also searched for line emission in addition to Fe K$\alpha$.
Reflection-dominated Seyfert 2 galaxies possess fluorescence emission 
lines due to elements such as Mg, Si, and S (e.g., Reynolds et al.\
1994).  We derive upper limits of EW$\sim 200$ eV for such features.

\subsection{Models That Include Thermal Emission} 

We also examine 
descriptions of the soft excess that include a Mewe-Kaastra (``Mekal'')
plasma model with solar abundances.  Describing the soft emission 
as purely thermal provides an acceptable fit for 
both the dual-absorber and reflection models (models 6 and 7).  
However, the best fit includes thermal and scattered emission
(model 8).  In this case, 
$\sim15\%$ of the soft excess is thermal with a temperature
of kT $=0.53^{+0.34}_{-0.26}$ keV, a $0.1-2$ keV flux of
$5.8^{+2.5}_{-2.3}\times10^{-14}$ ergs cm$^{-2}$ s$^{-1}$,
and a $0.1-2$ luminosity of
$5.9^{+2.5}_{-2.4}\times10^{39}$ ergs s$^{-1}$.
The properties of the thermal component are 
consistent with those of other elliptical galaxies with 
velocity dispersions similar to NGC 1052 (Davis \& White 1996).
This result, combined with the extended nature the soft X-ray emission 
in the HRI (Guainazzi \& Antonelli 1999), suggests that the
thermal component originates in the  
hot interstellar medium of the elliptical galaxy.  

The thermal component we measure can account 
for only 20\% of the observed luminosity in the HRI 
extent quoted by  
Guainazzi and Antonelli (1999).  However, there are many 
uncertainties in both measurements.  For example, the 
HRI data quality make is difficult to determine
exactly what fraction of the source is extended.  Also,
the HRI and PSPC have different sensitivities at 0.5 keV 
and so a direct comparison would have to take this into 
account.  In our best-fit model that includes scattered 
and thermal emission, the scattered fraction is 
higher than expected ($\sim11\%$), which suggests 
we may be underestimating 
the thermal contribution and overestimating the scattering
contribution.  One way this could happen is 
if we have the abundances wrong. 
Better data are needed to resolve this issue. 

\subsection{Summary of Spectral Fitting Results}

We have shown that the nuclear source in NGC 1052 
is heavily absorbed by gas having a column density of
$\ge10^{23}$ cm$^{-2}$.  Compton reflection models
imply that we see less than $5\%$ of the intrinsic source directly,
while dual-absorber models require $\sim70\%$ of the source to be 
absorbed by a column of $N_{\rm H} = 3\times10^{23}$ cm$^{-2}$
and $\sim30\%$ to be absorbed by a column of 
$N_{\rm H} = 3-5\times10^{22}$ cm$^{-2}$.  The weakness of the 
Fe K$\alpha$ line compared to that predicted if the X-rays  
are reflected argues in favor of a complex absorption 
structure (represented by the dual absorption 
model).  The prominent soft excess can be described as nuclear 
flux scattered by ionized gas or as partial covering of the 
nuclear source, plus a 15\% contribution from 
a $\sim0.5$ keV thermal component, which most likely 
arises from the galaxy.

\section{Discussion} 

As already mentioned, the observed $2-10$ keV luminosity 
of NGC 1052 is $\sim 4\times10^{41}$ ergs s$^{-1}$.  Our
most plausible physical models imply absorption-corrected 
luminosities of $\sim2$ to 50 times this value for the 
dual-absorber and reflection models, respectively.  Both 
of these models require gaseous columns of $N_{\rm H} \ga
3\times10^{23}$ cm$^{-1}$.  Such large absorbing columns
are entirely consistent with X-ray observations of other 
megamaser galaxies.  For example, NGC 1068 (Iwasawa et al.\
1997), NGC 4945 (Done, Madejski \& Smith 1996), and 
Circinus (Matt et al. 1996) are completely blocked in 
the $2-10$ keV band.

We can compare the density of the gas inferred from the 
maser observations with the column density inferred from the X-rays.
The size of the maser spots in NGC 1052 is estimated to 
be $\le10^{17}$ cm (Claussen et al.\ 1998).  The gas should
mase most efficiently at a density a bit below the quenching 
density.   Thus a reasonable estimate for the density 
would be $10^8-10^{10}$ cm$^{-3}$, which predicts a
column density less than $10^{25} - 10^{27}$ cm$^{-2}$.
This upper limit is entirely consistent with the 
inferred X-ray column of $>3\times10^{23}$ cm$^{-2}$.

The relationship between the X-ray absorber and the masing 
gas in NGC 1052 is different from the most well-studied 
megamaser galaxy, NGC 4258.  In NGC 4258, the masers and 
the absorption originate in an approximately Keplerian
accretion disk that orbits the central source.  In NGC 1052,
the VLBA map shows the maser spots to be at a projected 
distance of $\sim0.04$ pc from the believed nucleus
(Claussen et al.\ 1998), and they lie along the direction 
of the radio jet.  The maser emissions are 
only seen from those molecular clouds being hit by or 
located in front of the radio jet.  For example, the masing 
clouds could be located in the accretion disk between us and the 
jet if the SW radio jet (with which the masers coincide)
is the one pointed away from Earth.  Conversely,
the X-ray absorption arises from gas along our line of 
sight to the nucleus proper. If the nuclear X-ray source is 
coincident with the radio nucleus inferred from 43 GHz
observations (Vermeulen et al.\ 1998), the observed masing 
clouds (Claussen et al.\ 1998) must be spatially separate from 
those that absorb the X-ray emission, although the latter
might also be part of the accretion disk.

The absence of a strong Fe K$\alpha$ fluorescence line
with an equivalent width of $1-2$ keV  
argues against a purely reflected X-ray spectrum, and also
seems to fit the pattern seen in other LINERS by e.g.,
Terashima et al.\ (1997).  This may mean that there is no 
optically-thick inner accretion disk around the nuclei of 
these objects.
An attractive explanation is that the inner disks are 
actually ion tori (i.e., advection-dominated disks) as 
proposed originally by Chen \& Halpern (1989).
More high-sensitivity X-ray observations of masing galaxies 
are needed to provide reliable evidence for or
against advection-dominated accretion disks. 

\section{Conclusions}

The observed X-ray spectrum of the LINER galaxy NGC 1052
is unusually flat, $\Gamma \sim 0.2$, between the energies 
of 1.5 and 6 keV.  This is a signature of significant 
absorption which modifies the intrinsic spectrum.  The 
nuclear X-ray emission either traverses a complex 
distribution of absorbing material with column densities
of up to a few $\times10^{23}$ cm$^{-2}$, or is Compton 
reflected from gas with a column $\ga10^{24}$ cm$^{-2}$;
however, the moderate Fe K$\alpha$ EW of $\sim270$ eV
favors the complex-absorber model as the more likely 
physical scenario.  The inferred $2-10$ keV intrinsic
(absorption-corrected) luminosity is $\sim8\times10^{41}$
or $\sim2\times10^{43}$ ergs s$^{-1}$ for the complex 
absorber and Compton-reflection models, respectively.
The large column densities are consistent with the 
presence of luminous water vapor masers in NGC 1052;
however, the absorbing gas appears to be spatially 
separate from the masing gas in this galaxy.

This research has made us of the HEASARC database.  KAW 
acknowledges support from NASA long-term space 
astrophysics grant NAG 53504.  CAH acknowledges support
from NATO grant SA.5-2-05 (GRG.960086).  ASW thanks 
NASA and NSF for support under grants NAG 81027,
NAG 53393, and AST 9527289.

\vfil\eject

\begin{deluxetable}{lccccccc}
\tiny
\tablecolumns{8}
\tablewidth{0pc}
\tablecaption{$ASCA$ + PSPC Spectral Results$^a$}
\tablehead{
Model & N$_{H}^b$ & $f_c^c$ or kT$^d$ & $\Gamma_{int}^e$ &
$A_{\Gamma_{int}}^f$ & $A_{\Gamma_{scatt}}^g$ & 
$f_i^h$ or $A_{\rm MEKAL}^i$ &
$\chi^2$/dof$^j$  \\
 & [$\times10^{22}$] & & & [$\times10^{-4}$] & [$\times10^{-4}$]
 & & }
\startdata
1 & 0.03(f) & \nix & 0.24$\pm0.05$ & 0.95$_{-0.06}^{+0.07}$
  & \nix & \nix & 512.2/325 \nl
2 & 11.5$_{-2.6}^{+3.3}$ & 0.80$_{-0.10}^{+0.07}$ & 1.26$\pm0.12$ 
  & 8.3$\pm1.5$ & \nix & \nix & 354.4/323  \nl
3 & 12.8$_{-2.5}^{+2.9}$ & \nix & 1.28$\pm0.15$ 
  & 6.8$\pm1.0$ & 1.50$\pm$0.09 & \nix & 354.6/322 \nl 
3$_{\rm G}$ & 11.6$_{-2.7}^{+3.5}$ & \nix & 1.31$\pm0.18$ 
  & 6.7$\pm$1.6 & 1.5$\pm0.08$ & \nix & 334.2/322 \nl
3$_{\rm 2G}$ & 4.5$_{-2.7}^{+5.7}$ & \nix & 1.58$_{-0.28}^{+0.30}$ 
  & 2.9$_{-1.1}^{+1.6}$ & 1.5$\pm$0.1 & \nix & 315.1/318 \nl 
4 & $4.9_{-1.4}^{+2.0}$ & \nix & $1.65_{-0.25}^{+0.37}$
  & $4.5_{-1.6}^{+1.7}$ & $1.5\pm0.14$ & \nix & 314.4/320 \nl
  & 30.0$_{-11.6}^{+16.8}$ & & & 13.0$_{-3.4}^{+4.5}$ & &  \nl
5 & 0.03(f) & \nix & 1.7(f) & 173$\pm$10 & \nix & 0.007$\pm$0.002
  & 341.9/324  \nl
6 & 0.03(f) & 0.53$_{-0.26}^{+0.34}$ & 1.7(f) & $180\pm15$ & 0.18$\pm0.09$
  & 0.006$\pm0.002$ & 327.8/322 \nl
7 & $0.88_{-0.34}^{+0.32}$ & 0.53(f) & 1.7(f) & $3.2\pm0.5$ & \nix
  & $0.61\pm0.08$ & 321.9/320 \nl
  & 18.6$_{-4.1}^{+5.2}$ & & & 15.9$_{-1.8}^{+2.1}$ & & \nl
8 & $3.05_{-1.20}^{+1.48}$ & 0.53(f) & 1.7(f) & 4.0$_{-1.5}^{+1.6}$ & 1.1$\pm$0.6
  & $0.25\pm0.09$ & 301.1/320 \nl
  & 25.8$_{-11.0}^{+17.5}$ & & & 9.7$_{-4.7}^{+6.1}$ & & \nl
\enddata
\end{deluxetable}

\vfill\eject
\clearpage

Notes to Table 1:

$^a${Models are (1) uniformly-absorbed power law,
(2) partially covered power law, (3) absorbed power law plus 
scattered flux, (4) dual absorber plus scattered flux,
(5) Compton reflection, 
(6) Compton reflection plus MEKAL plasma, 
(7) dual absorber plus MEKAL plasma, 
(8) dual absorber plus scattered flux and MEKAL plasma.
All models include Galactic absorption with $N_{\rm H}=3\times10^{20}$
cm$^{-2}$. For reflection models, $\theta=60^{\circ}$ and
$\Omega=2\pi$.  The photon index of the scattered component
($\Gamma_{scatt}$) is assumed equal to the intrinsic photon 
index ($\Gamma_{int}$).  Models $3_{\rm G}$ through 6 include an 
Fe K$\alpha$ line; Model $3_{\rm 2G}$ contains an additional
Gaussian (Table 2).  A fixed parameter is denoted by (f).
Errors on normalization are $90\%$ confidence for 1 
free parameter ($\chi^2 + 2.7$).  Other errors are
90\% confidence for 2 (models 1, 3, 4), 3 (models 2,
$3_{\rm G}$, 5, 6, 7)
or 4 (model $3_{\rm 2G}$, 8) parameters, which is $\chi^2 + 4.6$, 
$\chi^2 + 6.4$, or $\chi^2 + 7.1$, respectively.

$^b$Column density in units of cm$^{-2}$.

$^c$Fraction of the X-ray source covered by 
   absorbing gas.

$^d$MEKAL plasma temperature in keV.

$^e$The intrinsic photon index.

$^f$Normalization of the intrinsic power law 
   in units of photons keV$^{-1}$ cm$^{-2}$ s$^{-1}$ at 1 keV.

$^g$Normalization of the scattered power law 
   in units of photons keV$^{-1}$ cm$^{-2}$ s$^{-1}$ at 1 keV.

$^h$Fraction of intrinsic nuclear flux that is 
   seen directly.

$^i$ Normalization of MEKAL plasma
   in units of photons keV$^{-1}$ cm$^{-2}$ s$^{-1}$ at 1 keV.

$^j$d.o.f=degrees of freedom, which is the number 
of data points minus the number of free parameters.

\vfill\eject
\clearpage 

\begin{deluxetable}{lcccccccc}
\tablecolumns{9}
\tablewidth{0pc}
\tablecaption{Spectral Results: Parameters of the Fe K$\alpha$ line$^{a}$}
\tablehead{
Model & E$_1^{b}$  &  $\sigma_1^{c}$  & Norm$_1^{d}$ & EW$_1^e$
  & E$_2^{b}$ & $\sigma_2^{c}$ & Norm$_2^{d}$ & EW$_2^e$\\
 & [keV] & [keV] & [$\times10^{-5}$] & [eV] &
[keV] & [keV] & [$\times10^{-5}$] & [keV] }
\startdata
$3_{\rm G}$  & $6.36^{+0.23}_{-0.16}$ & 0.02(f) & $1.78\pm1.04$
   & $302\pm190$ & \nix & \nix & \nix & \nix \nl
$3_{\rm 2G}$ & 6.37(f) & 0.02(f) & $1.25\pm0.56$ & $200\pm170$
   & $5.5^{+1.2}_{-4.0}$ & $2.4^{+3.7}_{-1.6}$ & $36\pm12$
   & $20\pm5$ \nl
4  & 6.37(f) & 0.02(f) & $1.9^{+0.9}_{-0.8}$ & $302^{+150}_{-140}$
   & \nix & \nix & \nix & \nix \nl
5  & 6.37(f) & 0.02(f) & $0.4^{+0.5}_{-0.4}$ & $46^{+58}_{-46}$
   & \nix & \nix & \nix & \nix \nl
6  & 6.37(f) & 0.02(f) & $0.3^{+0.4}_{-0.3}$ & $35^{+46}_{-35}$
   & \nix & \nix & \nix & \nix \nl
7  & 6.37(f) & 0.02(f) & $1.8\pm0.7$ & $267\pm100$
   & \nix & \nix & \nix & \nix \nl
8 & 6.37(f) & 0.02(f) & $1.8\pm0.8$ & $270\pm120$ 
   & \nix & \nix & \nix & \nix \nl
\enddata
\label{tab:feK_results}
\tablenotetext{a}{Models are defined in Table 1.
A fixed parameter is denoted by (f).}
\tablenotetext{b}{Gaussian energy.}
\tablenotetext{c}{Gaussian width.}
\tablenotetext{d}{Gaussian line flux in units 
of photons cm$^{-2}$ s$^{-1}$.}
\tablenotetext{e}{Gaussian equivalent width.}
\end{deluxetable}

\clearpage

\begin{figure}
\plotfiddle{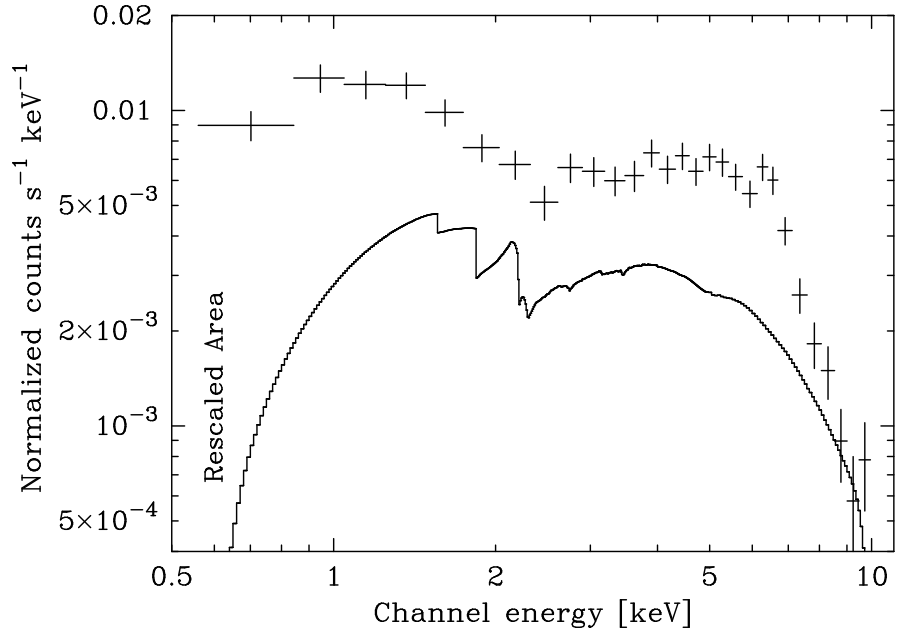}{150pt}{-90}{50}{50}{-200}{280}
\caption[ ]{
The $ASCA$ spectrum of NGC 1052.  The signal-to-noise ratio
is maximized by coarsely binning the averaged data from each
set of detectors and then plotting the SIS data below 5 keV
and the GIS data above 5 keV.  The solid line is the
corresponding telescope-plus-detector effective area
curve sampled at full spectral resolution and arbitrarily rescaled.
\label{fig:Figure1}
}
\end{figure}

\clearpage

\begin{figure}
\plotfiddle{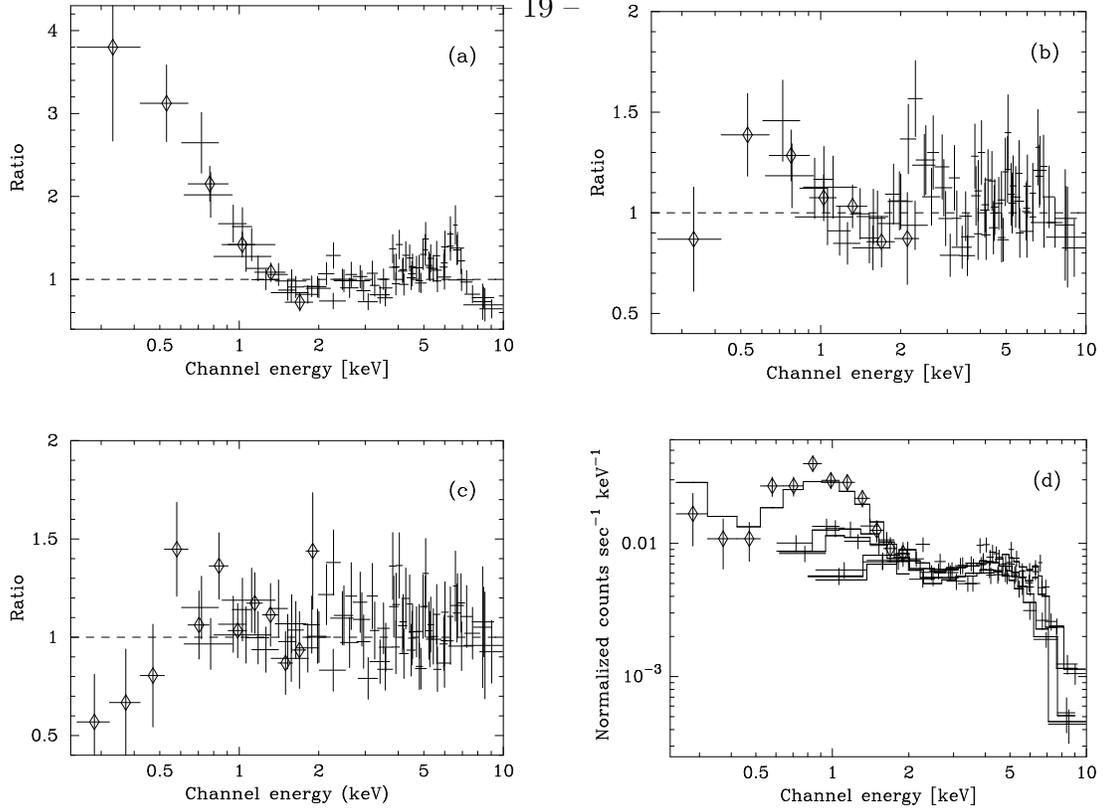}{150pt}{0}{60}{60}{-200}{10}
\caption[ ]{
The $ASCA$ SIS and GIS data (crosses) and $Rosat$ PSPC data
(diamonds).  The first three panels show 
the ratio of the data to a model that consists of (a) a
power law with Galactic absorption (Table 1, model 1);
(b) an absorbed power law above $\sim2$ keV, a scattered
power law absorbed by only the Galactic column below
$\sim2$ keV and a narrow Gaussian at 6.37 keV (Table 1,
model $3_{\rm G}$); (c) a dual absorber plus scattered
flux model (Table 1, model 4).  Panel (d) shows the        
observed spectrum and the dual-absorber plus scattered
flux model (panel c) folded through the instrumental
response.  Data from all 5 instruments are coarsely binned and 
plotted separately.  These diagrams   
illustrate the level of goodness of the statistics 
and so are not optimized for discrimination between the
SIS and GIS data nor between observed and modeled data.
The PSPC data in panels c and d are binned less coarsely 
than in panels a and b to better show the deviations
near 0.5 keV caused by the lack of including a thermal 
component in model 4.
\label{fig:Figure2}
}
\end{figure}

\clearpage 

\begin{figure}
\plotfiddle{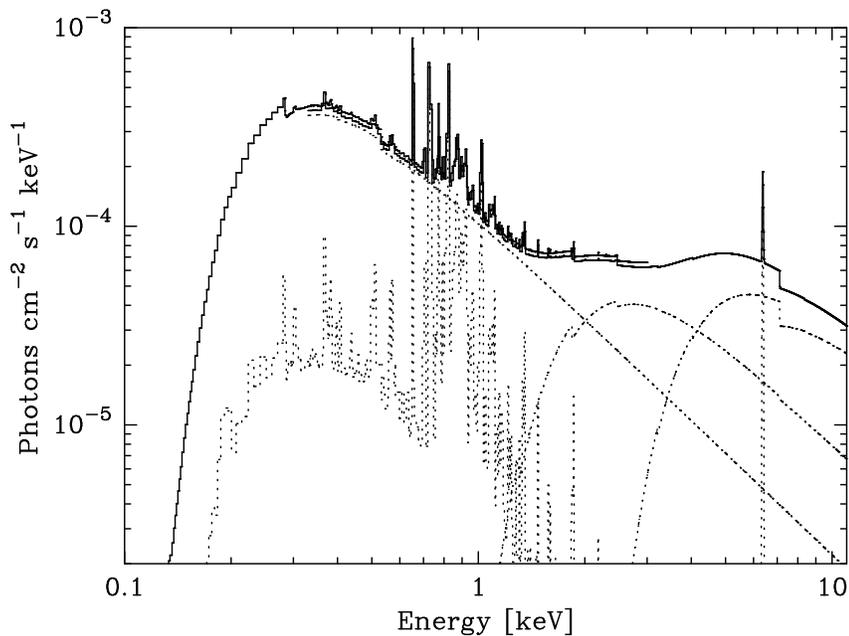}{150pt}{-90}{50}{50}{-200}{280}
\caption[ ]{
The model that best describes the $\sim0.1-10.0$ keV spectrum 
of NGC 1052: a power-law continuum 
with $70\%$ absorbed by $N_{\rm H1}=3\times10^{23}$ cm$^{-2}$
and $30\%$ absorbed by $N_{\rm H2}=3\times10^{22}$ cm$^{-2}$,
a soft X-ray component that consists of 85\%
scattered flux and 15\% thermal plasma emission with kT = 0.53 keV, 
and an unresolved Fe K$\alpha$ line with EW = 270 eV. 
\label{fig:Figure3}
}
\end{figure}

\end{document}